\documentclass[preprint]{aastex}

\slugcomment{Accepted to ApJ}

\shorttitle{Chemical Evolution in Oscillatory Star Formation}
\shortauthors{Hirashita, Burkert, \& Takeuchi}

\def\SFR{{\rm SFR}}
\def\oxy{{\rm O}}
\def\fe{{\rm Fe}}
\def\ins{{\rm ins}}
\def\del{{\rm del}}

\begin{document}


\title{CHEMICAL EVOLUTION OF THE GALAXY BASED ON THE OSCILLATORY
STAR FORMATION HISTORY}

\author{\bf Hiroyuki Hirashita\altaffilmark{1}
}
\affil{Department of Astronomy, Faculty of Science, Kyoto University,
Sakyo-ku, Kyoto 606-8502, Japan}
\email{hirasita@kusastro.kyoto-u.ac.jp}

\author{\bf Andreas Burkert}
\affil{Max-Planck-Institut f\"{u}r Astronomie, K\"{o}nigstuhl 17,
D-69117 Heidelberg, Germany}
\email{burkert@mpia-hd.mpg.de}

\and

\author{\bf Tsutomu T. Takeuchi}
\affil{Division of Particle and Astrophysical Sciences,
Nagoya University, Chikusa-ku, Nagoya 464-8602, Japan}
\email{takeuchi@u.phys.nagoya-u.ac.jp}


\altaffiltext{1}{Research Fellow of the Japan Society for the Promotion of
Science.}


\begin{abstract}

We model the star formation history (SFH) and the chemical evolution
of the Galactic disk 
by combining an infall model and a limit-cycle model of the
interstellar medium (ISM). Recent observations have shown that the SFH
of the Galactic disk violently variates or oscillates. We model
the oscillatory SFH
based on the limit-cycle behavior of the fractional masses
of three components of the ISM.
The observed period of the oscillation ($\sim 1$ Gyr) is
reproduced within the natural parameter range.
This means that we can interpret the oscillatory SFH
as the limit-cycle behavior of the ISM. We then test the
chemical evolution of stars and gas in the framework of the
limit-cycle model, since the oscillatory behavior of the SFH
may cause an oscillatory evolution of the metallicity.
We find however that the oscillatory behavior of metallicity is not
prominent because the metallicity reflects the past integrated SFH.
This indicates that the metallicity cannot be used to distinguish an
oscillatory SFH from one without oscillations.

\end{abstract}




\section{INTRODUCTION}\label{sec:intro}

Revealing the star formation histories (SFHs) of galaxies is
essential in understanding the galaxy formation and evolution.
The SFH of the Galaxy (Milky Way) is worth studying, since a large
number of stars are observed individually and the SFH is
inferred directly from the age distribution of the stars. The SFH is
closely related to the
chemical evolution of the Galaxy. For example,
the age--metallicity
relation (e.g., \citealt{pagel97}) of Galactic stars is generally
believed to originate from the chemical enrichment of the
Galaxy as a result of star formation.

\citet{eggen62} have pioneered the modeling of the Galactic SFH and its
chemical evolution.
{}From the correlation between the ultraviolet excess and orbital
eccentricity of stars, they have concluded that the Galaxy formed
by collapse on a free-fall timescale from a single protogalactic
cloud. An alternative picture of halo formation has been proposed
by \citet{searle78}. They have argued that the Galactic system
formed from the capture of fragments such as dwarf galaxies
over a longer timescale than that proposed by Eggen et al.
In any case, determining the timescale of the infall of matter and
the chemical enrichment is an important issue to resolve 
the formation history of the Galaxy.

A number of papers have investigated
the formation (e.g., \citealt{burkert92}) and chemical evolution
(e.g., \citealt{matteucci89}) of
the Galaxy and other spiral galaxies (e.g., \citealt{LB75,sommer96}).
Many models of the SFH of the Galaxy have
treated the formation of the Galactic disk through gas infall
from the halo. This
scenario (the so-called infall model) can be consistent with the
age--metallicity relation of the disk stars
(e.g., \citealt{twarog80}), if a reasonable SFH is used. Moreover, 
the infall model provides a physically reasonable way of solving
the G-dwarf problem (e.g., \citealt{pagel97}, p.236), contrary to
the closed-box model which tends to overpredict the number of the
low-metallicity stars.

Since stars are formed from interstellar medium (ISM), 
one of the factors that determine the star formation rate (SFR)
is the gas content of galaxies. 
Indeed, the SFR and the gas density is closely
related \citep{kennicutt98}. The most commonly used relation is
called the Schmidt law \citep{schmidt59}.
It assumes that $\SFR\propto\rho^n$,
where $\rho$ is the gas density and $n=1$--2.
As long as such a law is assumed and the infall of gas occurs
continuously as a smooth function of time, the predicted
SFH is also a smooth function of time.

Though the ``classical'' (i.e., {\it smooth}) infall model is widely
accepted, there are observational data that suggest 
{\it intermittent} or {\it oscillatory} star
formation activities in spiral galaxies. This means that the
SFH is not a smooth function of time. 
\citet{kennicutt94} have shown that the ratio of present-to-past SFR
in spiral sample has a significant scatter. More recently,
Tomita, Tomita \& Sait\={o} (1996) have analyzed the far-infrared to
$B$-band flux ratio $f_{\rm FIR}/f_B$ of 1681 spiral galaxies
(see also \citealt{devereux97}).
The indicator $f_{\rm FIR}/f_B$ represents the ratio between the
present SFR and the averaged SFR over the recent Gyr.
They have shown order-of-magnitude spread of $f_{\rm FIR}/f_B$ and
suggested a violent temporal variation of the SFR.

The intermittence of the SFH in the Galactic disk is recently suggested
by \citet[hereafter R00]{rocha00a}.
They have provided the SFH of the Galaxy inferred from the stellar age
of the solar neighborhood, using 552 late-type stars.
The age of each star has been estimated from the chromospheric
emission in the $\mbox{Ca}\,${\sc ii} H and K lines \citep{soderblom91}. 
After metallicity-dependent age correction, completeness correction, 
and scale-height correction\footnote{The scale height is dependent on 
the age of stars.}, they have derived the age distribution of the stars. 
Then, after correcting for evolved stars,
they have derived the SFH. They have also asserted that
their SFH derived from the stars in the solar neighborhood
is representative of the SFH in the whole disk,
since the diffusion timescale of stars is much shorter than
the Galactic age.
The discussion in this paper is based on Figure~2 of R00.
Based on their data, R00 have suggested
that the star formation activity of the disk is intermittent or 
variates violently. Their suggestion has been statistically
confirmed by Takeuchi \& Hirashita (2000, hereafter TH00), who have
also shown that the typical timescale of the variation
is 2 Gyr. We note that \cite{hernandez00} have also
found an oscillatory component of the SFH in the solar neighborhood.

Theoretically, 
the intermittent or oscillatory SFH is easily reproduced if
we treat the ISM as a nonlinear open system \citep{ikeuchi88}.
\citet[hereafter IT83]{ikeuchi83} 
have considered the ISM composed of three phases
(cold, warm, and hot) as suggested by \citet{mckee77} and modeled
the time evolution
of the fractional masses of the three components (see also
\citealt{habe81}).
Since the mass exchange among the three components
is a nonlinear process, the limit-cycle evolution of the fractional
masses can emerge (see also \citealt{scalo86,korchagin94}).
The limit-cycle behavior is supported by \citet{kamaya97},
who have interpreted the various levels of the star formation 
activities in spiral galaxies shown observationally by
Tomita et al.\ (1996) in the framework of the nonlinear open system
model. Their interpretation is based on the galaxy-wide limit-cycle
behavior of the ISM.

Another interesting topic is the chemical evolution in such a
limit-cycle ISM. If the oscillatory SFH is considered,
we may find an oscillation in a chemical enrichment process.
For example, the age--metallicity relation of the stars
in the Galactic disk may scatter because of the oscillation.
We will examine quantitatively such a scatter caused by
the limit-cycle evolution.

In this paper, we model the oscillatory SFH in the Galactic
disk proposed by R00
by combining the infall model and the limit-cycle model.
The chemical evolution in the oscillatory SFH is also
investigated.
This paper is organized as follows. First, in \S~\ref{sec:model}
we model the chemical evolution of the Galactic disk
by using the infall model. Then, in \S~\ref{sec:limit}
we review the limit-cycle model of ISM. Some results derived from
the equations are described in \S~\ref{sec:result}.
Finally, we discuss the SFH and the chemical evolution
in the limit-cycle ISM in \S~\ref{sec:discuss}.

\section{CHEMICAL EVOLUTION MODEL}\label{sec:model}

The chemical evolution model of the Galactic disk is constructed
here. The model is based on the infall model,
which is characterized by the gradual gas infall from the halo.
We adopt a one-zone model for simplicity. In other words,
the phenomena of the ISM are averaged in space.
This simple treatment is advantageous because the response of
the chemical evolution on the parameters is easy to examine.
When we compare the result with the observational data, however,
we should be careful whether the data are averaged or not.
We comment on the one-zone
approximation in \S~\ref{subsec:onezone}.

\subsection{Gas and Metal}

The changing rates of the gas mass ($M_{\rm g}$) and metal mass
($M_i$, where $i$
denotes the species of the metal;
$i=\fe $, O, etc.) in the Galactic disk are described by a 
set of differential equations
\begin{eqnarray}
\frac{dM_{\rm g}}{dt} & = & -\psi +E+F \, ,\label{eq:gas1} \\
\frac{dM_i}{dt} & = & -X_i\psi +E_i+FX_i^{\rm f} \, ,\label{eq:metal}
\end{eqnarray}
where $\psi$ is the SFR, $E$ is the
total injection rate of gas from stars, $F$ is the rate of gas
infall from
halo, $X_i$ is the abundance of $i$ (i.e., $X_i\equiv M_i/M_{\rm g}$),
$E_i$ is the injection rate of element $i$ from stars,
and $X_i^{\rm f}$ is the abundance of $i$ in the infall material
(see e.g., \citealt{tinsley80} for the basic treatment of chemical
evolution of galaxies). Introducing $X_i^{\rm f}$ enables us to
treat the infall of pre-enriched gas.
An early enrichment in the halo may be important for the initial
phase of the disk formation (e.g., \citealt{ikuta99}).

In this paper, we choose two tracers for the metallicity, O and Fe.
Almost all the oxygen is produced by high-mass stars, while
the iron is produced mainly by Type Ia supernovae (SNe) as well as
by high-mass
stars. Thus, we include the contribution from Type Ia SNe in our
formulation. We adopt the combination of the instantaneous recycling
approximation and the delayed production approximation as formulated
by Pagel \& Tautvai\v{s}ien\.{e} (1995). The evolution of
Fe abundance based on a model of Type Ia SN has been considered in
\citet{kobayashi98}. 
With these approximations, $E$ and $E_i$ at $t$ are expressed by using
the SFR as a function of time, $\psi (t)$:
\begin{eqnarray}
E & = & R_\ins \psi (t)+R_\del \psi (t-\tau ) \, ,\label{eq:e} \\
E_i & = & [R_\ins X_i(t)+Y_{i,\,\ins }]\psi (t)+
[R_\del X_i(t-\tau )+Y_{i,\,\del }]\psi (t-\tau ) \, ,\label{eq:ei}
\end{eqnarray}
where $R$ and $Y_i$ are the returned fraction of gas from stars
and the fractional mass of the newly formed element $i$, respectively,
and the subscripts ``ins'' and ``del'' denote the instantaneous
recycling and delayed production parts, respectively.
The delay time $\tau$ is set to 1.3 Gyr, according to \S~3 of
Pagel \& Tautvai\v{s}ien\.{e} (1995).

We assume that the disk begins to  form at $t=0$.  Its age is assumed to
be 15 Gyr.
When $t-\tau <0$, all the functions whose arguments depend on $(t-\tau)$
are set to zero; for example, $\psi (t-\tau )=0$ when
$t<\tau$. The initial condition is summarized in
\S~\ref{subsec:initial}.

With the above expressions (eqs.\ [\ref{eq:e}] and [\ref{eq:ei}]),
equation (\ref{eq:gas1}) becomes
\begin{eqnarray}
\frac{dM_{\rm g}}{dt}=-(1-R_\ins )\psi (t)+R_\del \psi
(t-\tau )+F(t) \, ,\label{eq:gas2}
\end{eqnarray}
and the combination of equations
(\ref{eq:metal}) and (\ref{eq:gas2}) leads to
\begin{eqnarray}
M_{\rm g}\frac{dX_i}{dt} & = & Y_{i,\,\ins }\psi (t)+Y_{i,\,\del }
\psi (t-\tau )+R_\del \psi (t-\tau )[X_i(t-\tau )-X_i(t)] \nonumber \\
& - &
F(t)[X_i(t)-X_i^{\rm f}] \, .\label{eq:metal2}
\end{eqnarray}

\subsection{Infall Rate of Gas}

For the infall rate $F$, we follow TH00.
They assumed an exponential form for the infall rate 
\begin{eqnarray}
F(t)=\frac{M_0}{t_{\rm in}}\exp (-t/t_{\rm in}) \, ,
\end{eqnarray}
where $M_0$ indicates the total mass that can fall into the
galaxy. In other words,
\begin{eqnarray}
\int_0^\infty F(t)\, dt=M_0 \, .
\end{eqnarray}
Normalizing equation (\ref{eq:gas2}) by $M_0$ leads to
\begin{eqnarray}
\frac{df_{\rm g}}{dt}=-(1-R_\ins )\tilde{\psi} (t)+R_\del 
\tilde{\psi}(t-\tau )+
\frac{1}{t_{\rm in}}\exp (-t/t_{\rm in}) \, ,\label{eq:normalgas}
\end{eqnarray}
where
\begin{eqnarray}
f_{\rm g}\equiv\frac{M_{\rm g}}{M_0}~~~\mbox{and}~~~
\tilde{\psi}\equiv\frac{\psi}{M_0} \, .
\end{eqnarray}
Equation (\ref{eq:metal2}) is also normalized by $M_0$ as
\begin{eqnarray}
f_{\rm g}\frac{dX_i}{dt} & = & Y_{i,\,\ins }\tilde{\psi}(t)+
Y_{i,\,\del }\tilde{\psi}(t-\tau )+R_\del \tilde{\psi}
(t-\tau )[X_i(t-\tau )-X_i(t)]\nonumber \\
& - &
\frac{X_i(t)-X_i^{\rm f}}{t_{\rm in}}\exp (-t/t_{\rm in}) \, .
\label{eq:normalmetal}
\end{eqnarray}

\subsection{Star Formation Law}

In order to include the three-phase model of the ISM composed
of cold, warm and hot components,
we modify the Schmidt law with the index $n=1$ (Schmidt 1959) as
\begin{eqnarray}
\psi =M_{\rm g}X_{\rm cold}/t_* \, ,\label{eq:schmidt}
\end{eqnarray}
where $X_{\rm cold}$ is the mass ratio of the cold component
to the total gas mass ($M_{\rm g}$) and $t_*$ is the timescale of
the cold-gas consumption to form stars. In other words, we consider
that stars are formed from cold clouds on a gas consumption
timescale of $t_*$. The time evolution of 
$X_{\rm cold}$ will be modeled based on IT83, which treated the
ISM as a nonlinear open system, in \S~\ref{sec:limit}.
Equation (\ref{eq:schmidt}) is equivalent to
\begin{eqnarray}
\tilde{\psi}=f_{\rm g}X_{\rm cold}/t_* \, .\label{eq:schmidt2}
\end{eqnarray}

\subsection{Choice of the Parameters of the Chemical Evolution}

According to \citet{pagel95}, we choose the fractional masses
of newly formed elements (see eq.\ [\ref{eq:ei}]) as follows:
$Y_{\rm O,\, ins}/X_{{\rm O}\odot}=0.70$,
$Y_{\rm O,\, del}/X_{{\rm O}\odot}=0.0$,
$Y_{\rm Fe,\, ins}/X_{\fe \odot}=0.28$, and
$Y_{\rm Fe ,\, del}/X_{\fe \odot}=0.42$, where the subscript
$\odot$ indicates the solar value. They explained the
observed metallicity of Galactic stars along with an infall
model. For the abundances in the inflow gas,
we examine two cases: one is the primordial case,
$(X_{\fe}^{\rm f}/X_{\fe \odot},\, X_{\oxy}^{\rm f}/X_{\oxy \odot})
=(0,\, 0)$; the other
is the pre-enriched case,
$(X_{\fe}^{\rm f}/X_{\fe \odot},\, X_{\oxy}^{\rm f}/X_{\oxy \odot})
=(0.1,\, 0.25)$\footnote{This means that ${\rm [Fe/H]}=-1$ and
${\rm [O/Fe]}=0.4$ in the inflow material.
 We note ${\rm [Fe/H]}=\log (X_\fe /X_{\fe\odot})$,
${\rm [O/H]}=\log (X_\oxy /X_{\oxy\odot})$, and
${\rm [O/Fe]}=\log (X_\oxy /X_{\oxy\odot})-\log (X_\fe /X_{\fe\odot})$.}.
The parameter in the delayed production approximation, $\tau$,
is set as $\tau =1.3$ Gyr.
The returned fractions of gas, $R_\ins $ and
$R_\del $, are determined as follows:
$R_\ins =0.16$, and
$R_\del =0.13$. The details about these values are described
in Appendix \ref{app:return}. We will determine $t_*$ and $t_{\rm in}$
in \S~\ref{subsec:param}.

\section{LIMIT-CYCLE MODEL OF THE ISM}\label{sec:limit}

We model the oscillatory behavior of the Galactic SFH proposed
by R00. IT83 have shown that an oscillatory behavior of the
fractional masses of three components (cold, warm, and hot)
emerges if one considers the ISM to be a nonlinear open
system. An introduction and details concerning nonlinear open
systems are
found in \citet{nicolis77}. We adopt the model by IT83 to explain
the oscillatory SFH by R00.

As long as the infall timescale ($t_{\rm in}$) is much longer than
the oscillatory timescale, the effect of the infall on the
limit-cycle evolution is not significant. Indeed, as
shown in Table \ref{tab1}, for the present case
the case where $t_{\rm in}\geq 9$ Gyr, which is much longer
than the period of oscillation proposed by R00 ($\sim 1$ Gyr;
see also \citealt{t2_00}).
Thus, we can apply the original model by IT83, which did not
include the effect of infall.

\begin{table}[htbp]
 \caption{Examined Parameters.\label{tab1}}
 \begin{tabular}{cccc}
  \tableline\tableline
   Model & $t_{\rm sf}$ (Gyr) & $t_*$ (Gyr) & $t_{\rm in}$ (Gyr) \\\tableline
   A & 6.0 & 3.9 & 23 \\
   B & 11  & 7.2 & 12 \\
   C & 15  & 9.8 &  9 \\
 \tableline
 \end{tabular}
\end{table}

\subsection{Model Equations}\label{subsec:model}

We review the model by IT83. This model is used to
calculate the time evolution of the
mass fraction of the three ISM phases (see also \S~1).
The result is used to calculate the SFR
through equation (\ref{eq:schmidt2}).

The ISM is assumed to consist of three
components (McKee \& Ostriker 1977); the hot rarefied gas
($T\sim 10^6$ K, $n\sim 10^{-3}$ cm$^{-3}$), the warm gas
($T\sim 10^4$ K, $n\sim 10^{-1}$ cm$^{-3}$), and cold
clouds ($T\sim 10^2$ K, $n\sim 10$ cm$^{-3}$).
The fractional masses of the three components are
$X_{\rm hot}$, $X_{\rm warm}$, and $X_{\rm cold}$, respectively.
A trivial relation holds:
\begin{eqnarray}
X_{\rm hot}+X_{\rm warm}+X_{\rm cold}=1 \, .\label{trivial}
\end{eqnarray}
The following three processes are considered
(see IT83 and \citealt{ikeuchi88} for the details):
[1] the sweeping of a warm gas into a cold component at the
rate of $aX_{\rm warm}$ ($a\sim 5\times 10^{-8}$ yr$^{-1}$);
[2] the evaporation of cold clouds
embedded in a hot gas at the rate of $bX_{\rm cold}X_{\rm hot}^2$
($b\sim 10^{-7}$--$10^{-8}$ yr$^{-1}$);
[3] the radiative cooling of a hot gas through collisions with
a warm gas at the rate of $cX_{\rm warm}X_{\rm hot}$
($c\sim 10^{-6}$--$10^{-7}$ yr$^{-1}$). Writing down
the rate equations and using equation (\ref{trivial}),
we obtain
\begin{eqnarray}
\frac{dX_{\rm cold}}{d\tau} & = & -BX_{\rm cold}X_{\rm hot}^2+
A(1-X_{\rm cold}-X_{\rm hot}) \, ,\label{eq:cold} \\
\frac{dX_{\rm hot}}{d\tau} & = & -X_{\rm hot}
(1-X_{\rm cold}-X_{\rm hot})+BX_{\rm cold}X_{\rm hot}^2 \, ,
\label{eq:hot}
\end{eqnarray}
where $\tau\equiv ct$, $A\equiv a/c$, and $B\equiv b/c$.

The solutions of equations (\ref{eq:cold}) and (\ref{eq:hot}) are
classified into the following three types, according to the
values of $A$ and $B$ (IT83):

1. $A>1$; all the orbits in the
$(X_{\rm cold},\; X_{\rm hot})$-plane
reduce to the node (0, 1),

2. $A<1$ and $B>B_{\rm cr}$; all the orbits reduce to a stable
focus $[(1-A)/(AB+1),\; A]$,

3. $A<1$ and $B<B_{\rm cr}$; all the orbits converge on a
limit-cycle orbit,

\noindent
where $B_{\rm cr}\equiv (1-2A)/A^2$. Apparently, case 3 is
important in the interpretation of the oscillatory SFH shown in
R00. Thus, we choose the parameters that satisfy case 3
as will be described in the next subsection.

\subsection{Choice of Parameters of the Limit-Cycle Model}
\label{subsec:param}

Since the timescale of the variation of SFR derived by R00
is $\sim 1$ Gyr
(see also TH00), we first investigate whether the
period of the limit-cycle can be the order of 1 Gyr. Indeed, a
Gyr-timescale cycle is possible in the natural parameter range.
According to Figure 3 of IT83, the period can be $\sim 10^2/c$
when we choose $A=0.3$ and $A=0.5$. Since $1/c$ is of the order
of $\sim 10^6$--$10^7$ yr, $10^2/c\sim 1$ Gyr is possible. Thus,
we choose $A=0.3$ or $A=0.5$ and $1/c=10^7$ yr to demonstrate
the Gyr-scale oscillation of the Galactic SFH.
The subsequent discussions are unchanged if we adopt another
set of the parameters that satisfies the oscillation period of
$\sim 1$ Gyr.

Here we confirm that the adopted parameters are within the
reasonable range of the physical properties of the ISM in the
Galactic disk. First, \citet{ikeuchi83} estimated $a$ from the
SN rate and the maximum radius of an SN remnant (SNR) and obtained
$a\simeq 5\times 10^{-8}$ yr$^{-1}$. Next, we estimate $b$ as the
reciprocal of the evaporation timescale of a cold cloud. The
evaporation timescale estimated in
\citet{hirashita00a} may be applicable in the present case
and we obtain $b\simeq 10^{-7}$ yr$^{-1}$. Finally, $c$ is
estimated from the collision rate of a cold cloud with SNRs.
The collision rate $t_{\rm col}$ is estimated as
$t_{\rm col}\simeq (\pi R_{\rm SNR}^2n_{\rm SNR}v)^{-1}$,
where $R_{\rm SNR}$ is the typical size of a SNR, $n_{\rm SNR}$
is the number density of SNRs in the interstellar space,
and $v$ is the typical relative velocity between a SNR
and a cloud. If we put $R_{\rm SNR}=50$ pc,
$n_{\rm SNR}=10^{-6}$ pc$^{-3}$,\footnote{The values of $R_{\rm SNR}$
and $n_{\rm SNR}$ are estimated according to \citet{ikeuchi83}.
We assume that the typical lifetime of a SNR is $10^7$ yr.}
and $v=100$ km s$^{-1}$, we obtain $t_{\rm col}\simeq 10^{7}$ yr.
This means that $c\simeq t_{\rm col}^{-1}\simeq 10^{-7}$ yr$^{-1}$.
$A=0.3$ and $B=0.5$ are easily satisfied if we assume for example
$a=6\times 10^{-8}$ yr, $b=10^{-7}$ yr, and $c=2\times 10^{-7}$ yr,
all of which are consistent with the above order-of-magnitude
estimates.

TH00 adopted a star formation law
$\tilde{\psi}=f_{\rm g}/t_{\rm sf}$ and did not consider
the effect of a multi-component medium  with phase changes. In order to
use their choices 
of the parameter values $A$ and $1/c$ we must relate their
$t_{\rm sf}$ (the timescale
in the classical {\it smooth} infall model) to our $t_*$
(the one in the {\it oscillatory} infall model).
This is achieved by averaging
the gas consumption rate defined as
$\tilde{\psi}/f_{\rm g}=X_{\rm cold}/t_*$
(eq.\ [\ref{eq:schmidt}]) over the whole galactic age where the
oscillatory part of the
efficiency is smoothed out and becomes $1/t_{\rm sf}$.
In other words, 
\begin{eqnarray}
\langle\tilde{\psi}/f_{\rm g}\rangle =
\langle X_{\rm cold}\rangle /t_*=1/t_{\rm sf} \, ,\label{eq:def_tsf}
\end{eqnarray}
where $\langle\cdot\rangle$ indicates the time average of the quantity
over the Galactic history ($0<t<15$ Gyr).
Since $X_{\rm cold}=0.65$ for
$(A,\, B)=(0.3,\, 0.5)$ (Appendix B),
$t_*=0.65t_{\rm sf}$. Following TH00, we examine the three sets of
$(t_{\rm sf},\, t_{\rm in})$ as summarized in Table \ref{tab1}.
They determined the parameters by fitting their infall model to
the observed SFH proposed by R00. All the three models provide
an almost identical SFH and reproduce the smoothed trend of
the SFH (T00). Thus, it is meaningful to examine all the three cases.
However, Model A gives an infall
timescale much larger than that in e.g., \citet{matteucci89},
although they assumed the same time dependence of infall as
that in this paper. This long timescale indicates that the
infall rate is almost constant over the history of the Galactic
disk. We note that Model A predicts the highest metallicity
of the three models (\S~\ref{sec:result}).

\subsection{Treatment of the Delayed Production}\label{subsec:delay}

Here we should comment on the delayed production approximation.
It assumes that all the delayed production at $t$ is determined by
the SFR at $t-\tau$. In reality, $\tau$ differs among Type Ia SNe.
Thus, the delayed production is determined by the averaged
SFR around $t-\tau$. Expecting that the lifetimes of Type Ia
progenitors are comparable to, or longer than, 1 Gyr
($\sim\mbox{the period of the
SFR oscillation in the Galaxy}$) the averaged SFR
around $t-\tau$ is described as
$M_{\rm g}(t-\tau )\langle X_{\rm cold}\rangle /t_*=
M_{\rm g}(t-\tau )/t_{\rm sf}$,
where $M_{\rm g}(t-\tau )$ is the gas mass at $t-\tau$ and
equation (\ref{eq:def_tsf}) is used.
Thus, we hereafter assume that
\begin{eqnarray}
\tilde{\psi}(t-\tau )=\frac{f_{\rm g}(t-\tau )}{t_{\rm sf}} \, .
\label{eq:delayedsfr}
\end{eqnarray}

\subsection{Comment on One-Zone Treatment}\label{subsec:onezone}

Following IT83, the structure of a model galaxy is
approximated by one zone.
The simplicity of the one-zone approximation gives the advantage 
that the background physical processes are easy to see. In this
subsection, we discuss the one-zone treatment.

Habe et al.\ (1981) stated in their \S~7 that for the one-zone
assumption to be acceptable it is necessary that the mean distance 
between supernova remnants (SNRs) be less than 100 pc
if a characteristic lifetime of SNRs of 
$\tau_{\rm life}\sim 10^7$ yr and a mean
expansion velocity of 10 km s$^{-1}$ are adopted. This is because
the SNRs should affect the whole disk for the one-zone treatment.
A distance of less than 100 pc means that there are $N\sim 10^4$
SNRs in a galaxy disk, if the disk size of 10 kpc is assumed.
This number is reasonable if SNe occur every $10^3$ yr
($\tau_{\rm life}/N\sim 10^7$ [yr]/$10^4$). Considering that the SN
rate in a spiral galaxy is typically 1/100--1/50 yr$^{-1}$
(Cappellaro et al.\ 1993), the mean distance between
SNRs is less than 100 pc even if 10--20 massive stars are clustered
in a region.

The information of a place can travel over
a distance of $c_{\rm s,\, eff}t_{\rm cyc}$ in one period of the
limit cycle, where $c_{\rm s,\, eff}$ is the effective sound speed
in the ISM and $t_{\rm cyc}$ is the oscillatory period.
Estimating these quantities as $c_{\rm s,\, eff}\sim 10$
km s$^{-1}$ (a typical velocity dispersion 
in the interstellar space) and $t_{\rm cyc}\sim 1$ Gyr, we obtain
$c_{\rm s,\, eff}t_{\rm cyc}=10$ kpc. Thus, the information of
the limit-cycle behavior can propagate over the whole disk.
Thus, the assumption of the limit-cycle behavior over the whole
disk may be good. In this paper, as a first
step, we treat the model galaxy as being a one-zone object. Here
we should note that the gas transport in the radial direction is
difficult if we consider the angular momentum conservation.
This difficulty may be a cause of the radial gradient of
the metallicity and the gas-to-star fraction in spiral galaxies.

The above discussions are not a satisfactory ``proof'' for the
limit-cycle oscillation on the scale of the whole Galactic disk.
(But it is important that it is not rejected.)
At present, thus, the cyclic star formation over the whole disk
is an assumption that easily explains the variation of the
star formation activity observed in the solar neighborhood by
R00. We note that
it also explains the variety of the star formation activity
of spiral galaxies (\citealt{kamaya97}). Hence, in this paper,
we base our discussion on the limit-cycle behavior on a whole-disk
scale.

When we compare our result with the observational data, we should
carefully consider to what extent the data is averaged.
The range of Galactocentric radii that enter in the average
depends on the age. Considering that the diffusion of stellar
orbits on a scale of 1 kpc occurs in 0.2 Gyr \citep{wielen77},
it is reasonable to assume that the observational quantities
are averaged on a scale of more than 1 kpc. Thus, as a first
step, we adopt the one-zone treatment for the Galactic disk
to see the chemical evolution in an oscillatory SFH.
We should extend our model to multi-zone treatment as
\citet{chiappini97} (see also \citealt{romano00} for a recent
work) in the future. Observationally, the formalism in
\citet{meusinger91} may be useful in order to link the global
observed SFH with the SFH in different Galactocentric annuli.

\subsection{Initial Conditions}\label{subsec:initial}

The initial condition is set as follows:
$X_\oxy (t=0)=0$, $X_\fe (t=0)=0$, $f_{\rm g}(t=0)=0$,
$X_{\rm cold}(t=0)=0.1$, and $X_{\rm hot}(t=0)=0.7$. The convergence
to the limit-cycle occurs on the timescale of a few
periods. The results are however not strongly dependent on the choice of the
initial conditions for $X_{\rm cold}$ and $X_{\rm hot}$.

\section{RESULTS}\label{sec:result}

In this section, the results calculated from the equations
above are presented. They are compared with the observational data.
Before displaying the results, we review the solving processes of
the equations. First,
the mass fraction of the cold gas, $X_{\rm cold}$, is
calculated by equations (\ref{eq:cold}) and (\ref{eq:hot}).
$X_{\rm cold}$ is used to determine the SFR at $t$ through
equation (\ref{eq:schmidt2}).
As for the delayed contribution expressed in $\tilde{\psi} (t-\tau )$,
we take into account the scatter of the lifetimes of progenitors
of Type Ia SNe (\S~\ref{subsec:delay}) and assume equation
(\ref{eq:delayedsfr}).
The SFH and the chemical evolution are modeled
by the infall model. The evolution of the gas mass normalized by
the total available gas mass $M_0$ is calculated by
equations (\ref{eq:normalgas}).
For $t_{\rm sf}$ and $t_{\rm in}$, we examine three cases listed
in Table \ref{tab1}. These three cases are also examined in TH00.
The chemical evolution is calculated by equation
(\ref{eq:normalmetal}).

\subsection{Star Formation History}

\begin{figure*}[htbp]
\figurenum{1}
\epsscale{1.1}
\plottwo{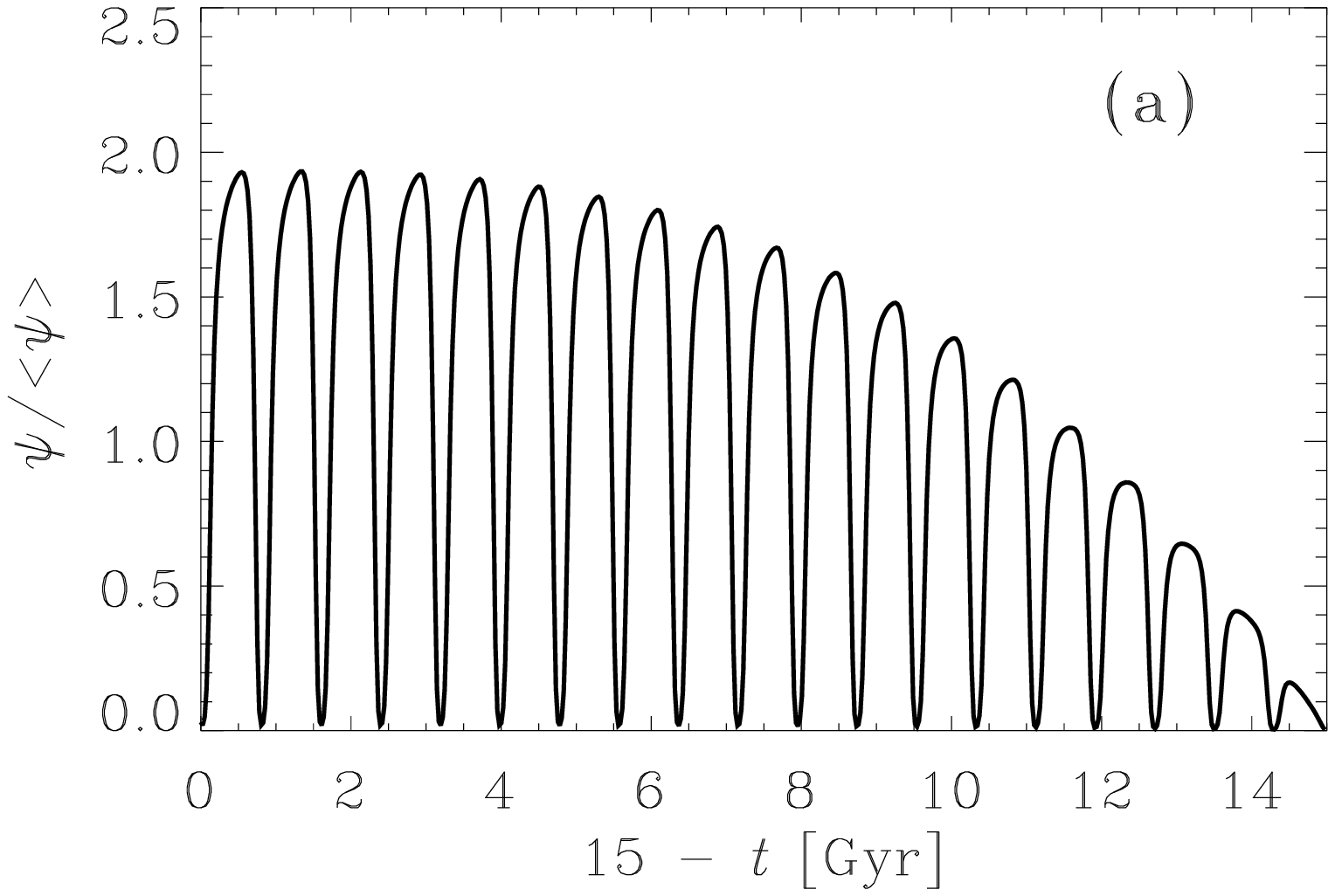}{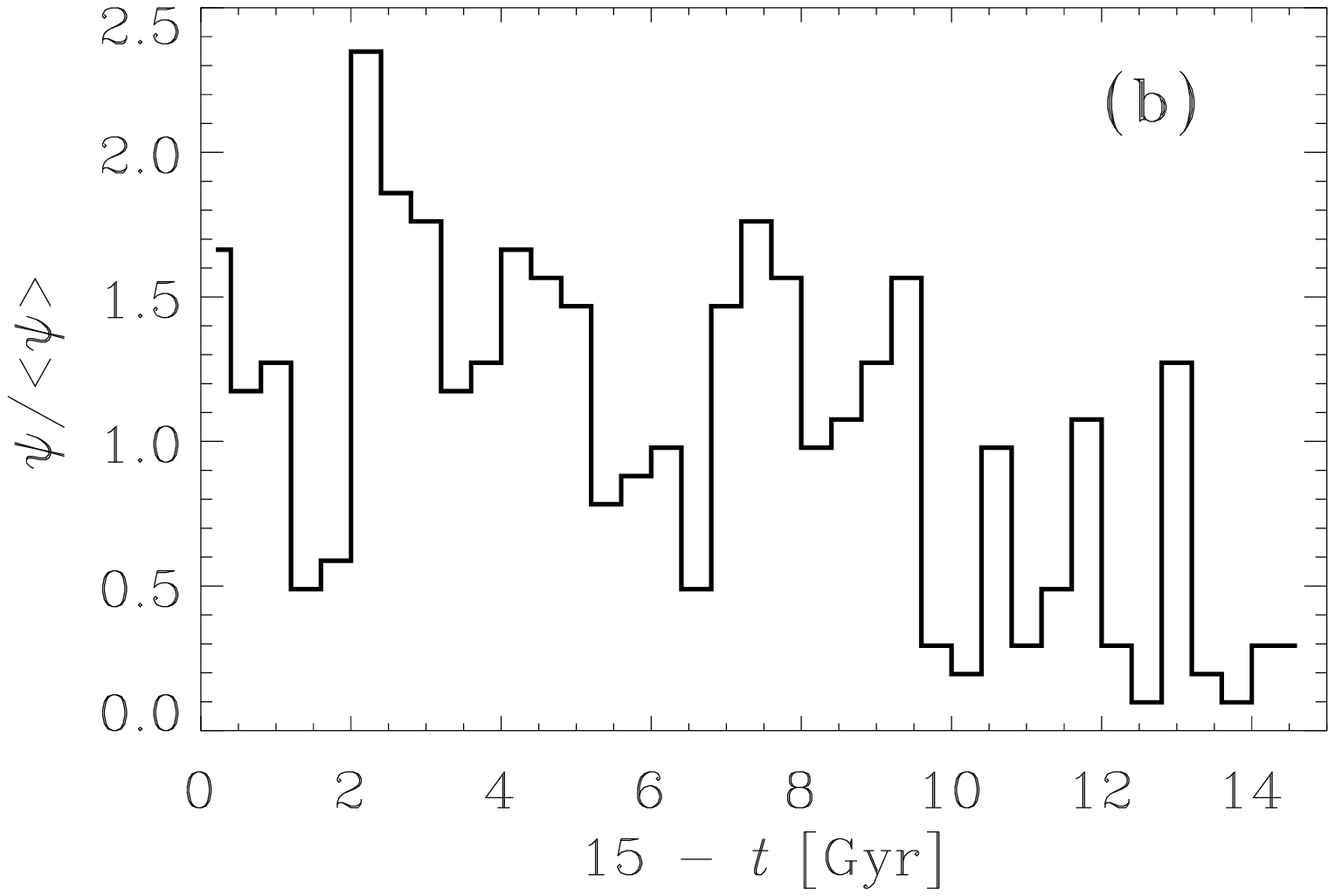}
\figcaption{({\it a}) Simulated star formation
history based on our model. The star formation rate ($\psi$)
as a function of look-back time [$(15-t)$ Gyr] is normalized with
the time-averaged value of $\psi$ ($\langle\psi\rangle$).
Since the three models shown in
Table \ref{tab1} result in almost an identical star formation
history, only Model A is shown.
 ({\it b}) Star formation history derived observationally
by \citet{rocha00a}.
\label{fig1}}
\end{figure*}

The SFH calculated by our model is
presented in Figure \ref{fig1}{\it a}. Since the 
three models predict almost the same SFH as
indicated by TH00, we present only the result of Model A
in Table \ref{tab1}. We also show
the SFH observationally determined by R00 in
Figure \ref{fig1}{\it b} in order to demonstrate the
qualitative similarity between the model prediction and the
observation.

\subsection{Metallicity Evolution}

We test the model with the metallicity data of Galactic
stars. Age--metallicity relation, G-dwarf metallicity
distribution, and [Fe/O]--[Fe/H] relation are examined.
First of all, we should note that the yields may be uncertain
because of the treatment of convection, nuclear reaction rates,
mass loss in the asymptotic giant branch phase, etc.
If the yields are systematically larger/smaller than
assumed in this paper, the metallicities predicted by
our model should be systematically larger/smaller.
Thus, quantitative agreement by fine tuning of the parameters
might be meaningless.
However, the qualitative behavior of the metallicity evolution
in the limit-cycle ISM is not altered even if the yield
changes.

\subsubsection{Age--metallicity relation}\label{subsec:amr}

The age--metallicity relation of stars in the Galactic disk
provides us with information
on its chemical enrichment history. Thus, our model
is worth testing by using the age--metallicity relation
of the stars in the solar neighborhood. The sample is provided by
\citet{rocha00b}, which used the same sample as R00.

\begin{figure*}[htbp]
\figurenum{2}
\epsscale{1.1}
\plottwo{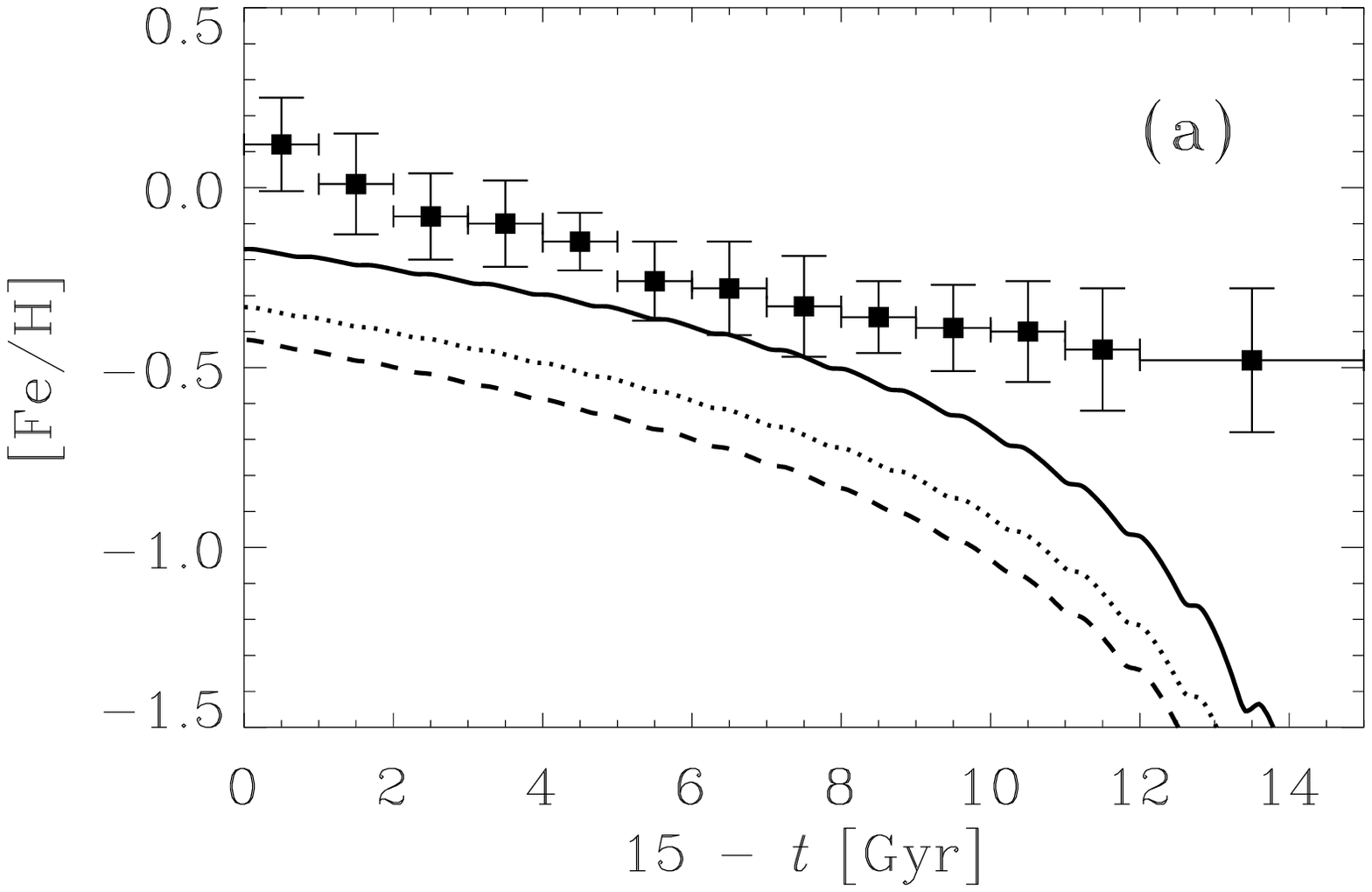}{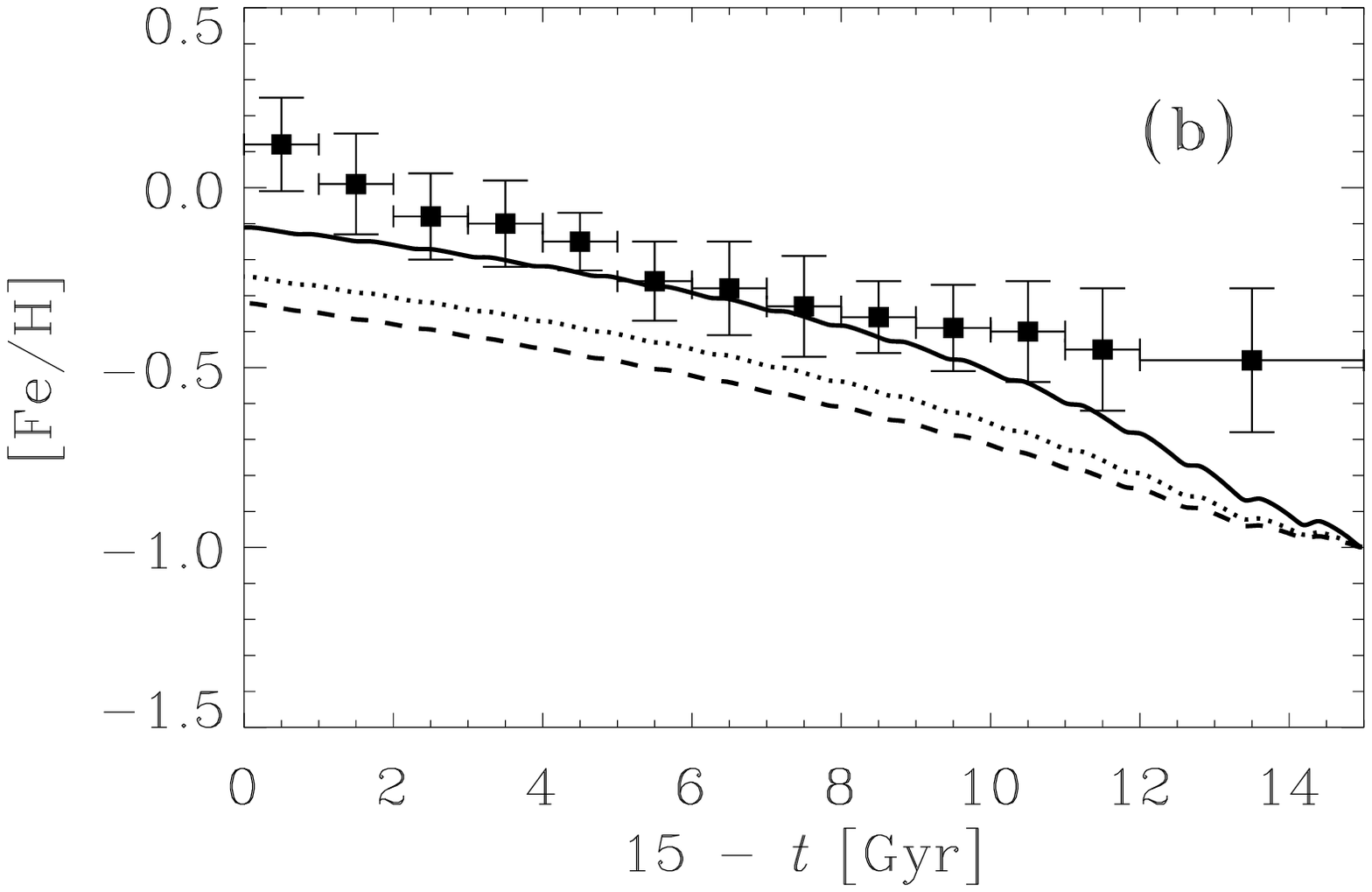}
\figcaption{Age--metallicity relation of the stars
in the solar neighborhood. [Fe/H] is used as an indicator for the
metallicity. The solid, dotted, and dashed lines
represent the results in Models A, B, and C, respectively.
The squares with error bars indicate the observational data point by
\citet{rocha00b}. ({\it a}) Abundance in the infalling gas is
assumed as ({\it a})
$(X_{\fe}^{\rm f}/X_{\fe \odot},\, X_{\oxy}^{\rm f}/X_{\oxy \odot})
=(0,\, 0)$ and ({\it b})
$(X_{\fe}^{\rm f}/X_{\fe \odot},\, X_{\oxy}^{\rm f}/X_{\oxy \odot})
=(0.1,\, 0.25)$.
\label{fig2}}
\end{figure*}

First, we examine the case where the infall gas is of primordial
abundance
[i.e., ($X_{\fe}^{\rm f}/X_{\fe \odot},$ $X_{\oxy}^{\rm f}/X_{\oxy \odot})
=(0,\, 0)$].
The age--metallicity relation predicted by our model is shown in
Figure \ref{fig2}{\it a}. The solid, dotted, and dashed lines represent
the results in Models A, B, and C, respectively.
We also present the observational data of the age--metallicity relation
by \citet{rocha00b} (see their Table 3).
Model A predicts the highest present metallicity, since its short gas
consumption timescale leads to the most efficient chemical enrichment.
Even in Model A, however, the discrepancy between
the model prediction and data is significant in the low-metallicity
range. \citet{rocha00b} noted the high initial metallicity
of the disk and attributed it to the pre-enrichment of the gas
before the formation of the first stars in the disk.
They also have shown that the age--metallicity relation determined
from the chromospheric age can deviate upward (i.e., metallicity is
overestimated) for larger ages from the
real relation because of the uncertainty in the age estimation.
Thus, we also examine the case where the infalling gas in enriched
with metal
[$(X_{\fe}^{\rm f}/X_{\fe \odot},\, X_{\oxy}^{\rm f}/X_{\oxy \odot})
=(0.1,\, 0.25)$]. The result is shown in Figure \ref{fig2}{\it b}.
We see that the discrepancy between the model prediction and
the observational data is reduced. 
{}From the viewpoint of modeling, the yield is also uncertain.
Thus, we do not try any fine-tuning the age--metallicity relation.

In spite of such uncertainty, we can discuss the qualitative
behavior of the age--metallicity relation. The amplitude of
the oscillation of
metallicity is smaller than the typical scatter of the observed data
points ($\sim 0.3$ dex).
The small amplitude is natural, because metallicity is determined by
all the past history of star formation, and thus the
present oscillatory star formation does not significantly contribute
to the metallicity.

Rocha-Pinto et al.'s data shown in Figure \ref{fig2}
seem to show a recent increase in [Fe/H]. The pre-enriched infall
cannot solve this increase, since the infalling gas has too low
a metallicity. However, we should carefully examine whether
the most recent data point in Figure \ref{fig2} represents the
SFH of the whole Galactic disk, because if the recent
chemical enrichment rate in the
solar neighborhood is significantly higher than that in
the whole Galaxy, the most recent data point would naturally  show a
systematically higher metallicity. Moreover, data sets shown
by other authors do not necessarily show such an increase in
the recent metallicity (e.g., \citealt{twarog80}). It
is necessary to analyze the observational data further
before we construct a theoretical model for the recent
[$15-t\la 1$ (Gyr)] increase in metallicity.

\begin{figure*}[htbp]
\figurenum{3}
\epsscale{0.65}
\plotone{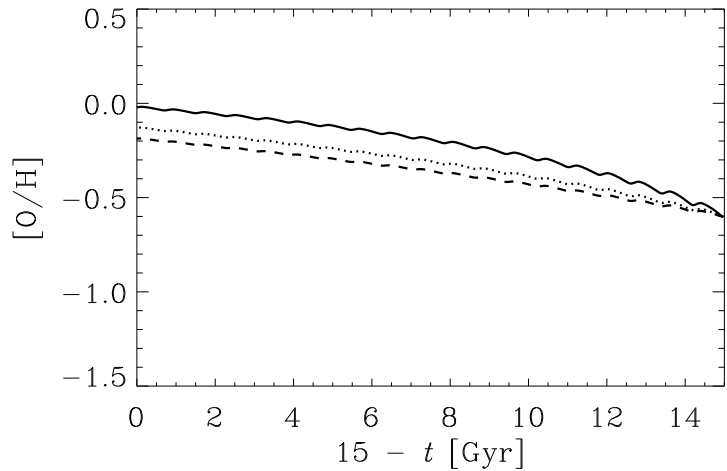}
\figcaption{Age--metallicity relation of the stars
in the solar neighborhood. [O/H] is used as an indicator for the
metallicity. The solid, dotted, and dashed lines
represent the results in Models A, B, and C, respectively.
For the initial enrichment,
$(X_{\fe}^{\rm f}/X_{\fe \odot},\, X_{\oxy}^{\rm f}/X_{\oxy \odot})
=(0.1,\, 0.25)$ is assumed.
\label{fig3}}
\end{figure*}

We can expect that the oscillation behavior is
more prominent for oxygen than for iron, because all the
oxygen is produced from the ``instantaneous'' part
(i.e., $Y_{\rm O,\, ins}\gg Y_{\rm O,\, del}$).
We present the result for the case of pre-enriched infall
[i.e., $(X_{\fe}^{\rm f}/X_{\fe \odot},\, X_{\oxy}^{\rm f}/X_{\oxy \odot})
=(0.1,\, 0.25)$] in Figure \ref{fig3}. Indeed, the amplitude of
the oscillation is larger than Figure \ref{fig2}{\it b}.
However, the oscillation would not explain the scatter
of the oxygen abundance of the stars in the Galactic disk.

\begin{figure*}
\figurenum{4}
\epsscale{1.0}
\plottwo{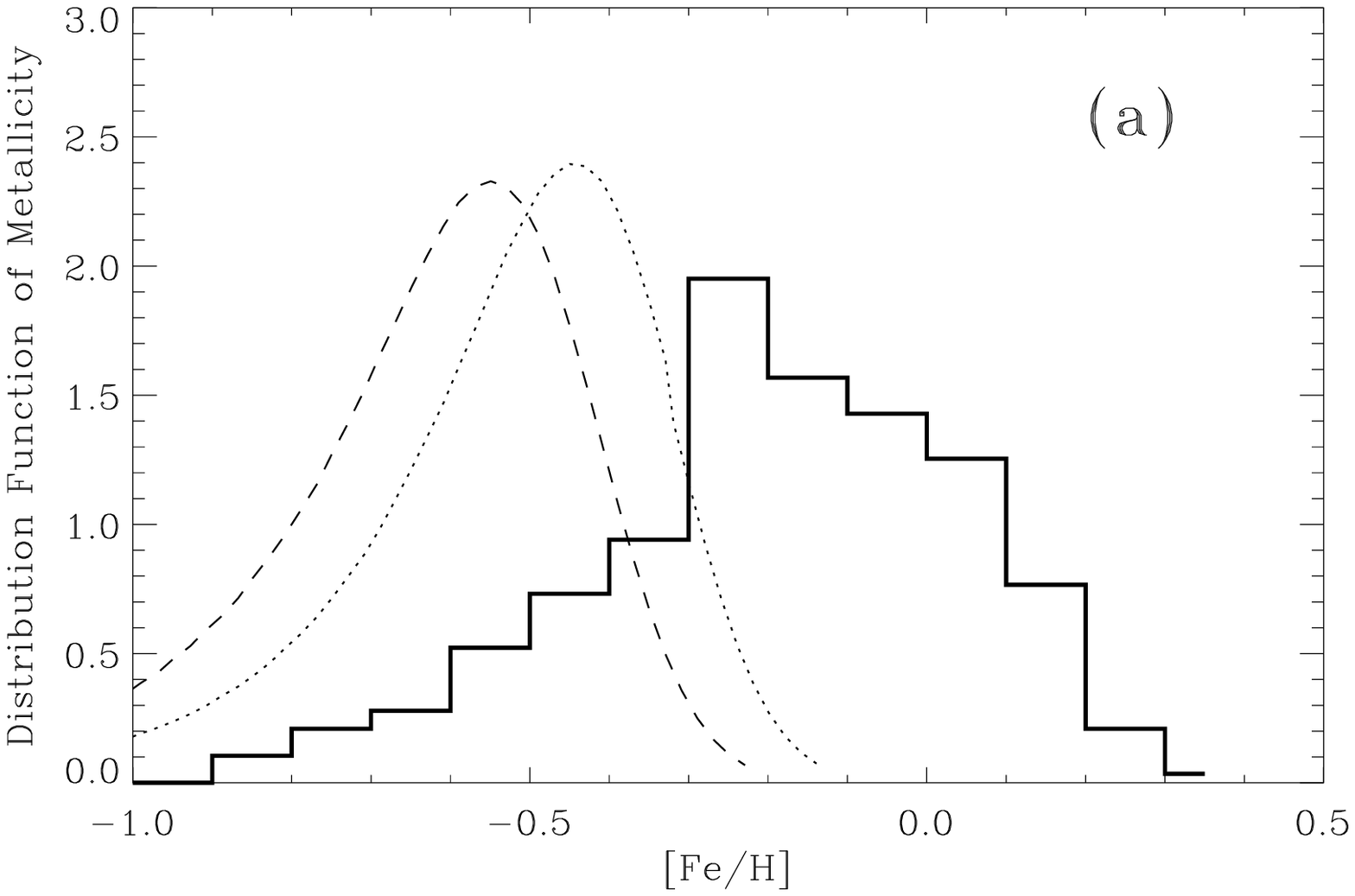}{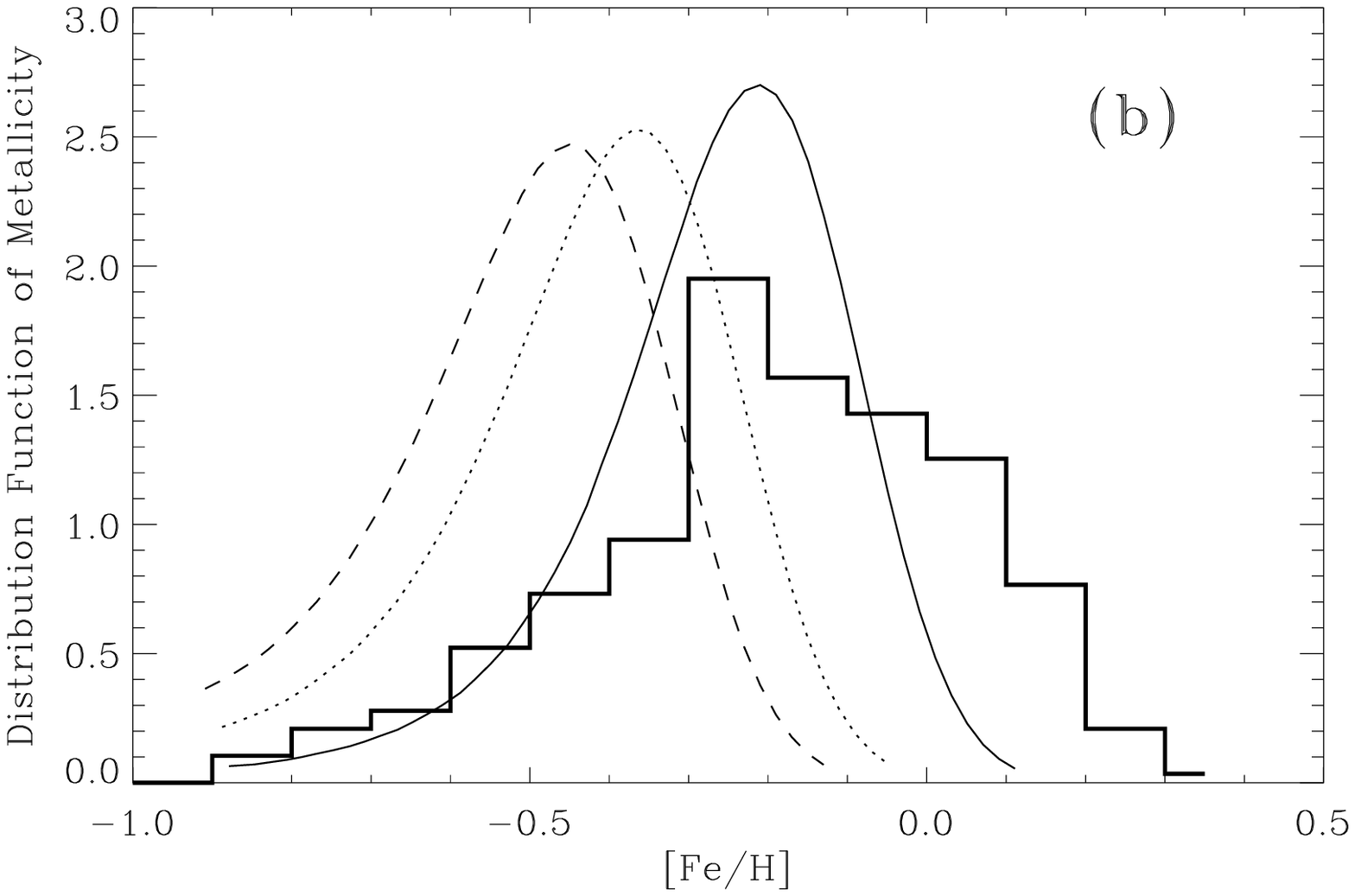}
\figcaption{Distribution of the G-dwarf metallicity.
The histogram shows the data by \citet{rocha96}.
The solid, dotted, and dashed lines represent the results in
Models A, B, and C, respectively.
All the distributions are normalized to unity when they are
integrated in the whole range of [Fe/H].
 ({\it a}) Abundance in the infalling gas is
assumed as ({\it a})
$(X_{\fe}^{\rm f}/X_{\fe \odot},\, X_{\oxy}^{\rm f}/X_{\oxy \odot})
=(0,\, 0)$ and ({\it b})
$(X_{\fe}^{\rm f}/X_{\fe \odot},\, X_{\oxy}^{\rm f}/X_{\oxy \odot})
=(0.1,\, 0.25)$.\label{fig4}}
\end{figure*}

\subsubsection{Metallicity distribution}

The G-dwarf metallicity distribution is also tested along with our
model, since the primary motivation for the infall model is
to solve the G-dwarf problem (e.g., \citealt{pagel97}).
The probability distribution function $P(\log X_i)$ of the
metallicity is calculated from our model as
\begin{eqnarray}
  P(\log X_i)\, d\log X_i=C\tilde{\psi}\, dt \, ,\label{prob_X}
\end{eqnarray}
where the constant $C$ is the normalization so that
\begin{eqnarray}
C\int_{0}^{15~{\rm Gyr}}\tilde{\psi}\, dt=1.
\end{eqnarray}
{}From equation (\ref{prob_X}), we obtain
the following analytical expression for $P$:
\begin{eqnarray}
  P(\log X_i) = C(\ln 10)\; \tilde{\psi} (t)X_i(t)\left(
  \frac{dX_i}{dt}\right)^{-1} \;,
\end{eqnarray}
where $dX_i/dt$ is calculated from equation (\ref{eq:normalmetal}).
In comparing the distribution function with the observational
data, we should take into account the scatter of the data. 
Here, we simply convolve $P$ with a Gaussian kernel as
\begin{eqnarray}
  P_{\rm conv}(\log X_\oxy )\equiv\int_{-\infty}^\infty P(u)\,
  \frac{1}{\sqrt{2\pi}\,\sigma}\,
  \exp\left[ -\frac{(\log X_\oxy -u)^2}{2\sigma^2}\right]\, d u \, ,
\end{eqnarray}
where we adopt $\sigma =0.1$ to compare with \citet{rocha96}.
We adopt these data because we would like to use a sample of
G-dwarfs, whose lifetimes are comparable to the age of the universe.
In Figure~\ref{fig4}, we show $P_{\rm conv}$ as a function of
[Fe/H]. The solid, dotted, and dashed lines represent the result
in Models A, B, and C, respectively. The histogram shows the
data by \citet{rocha96}. The two figures ({\it a}) and
({\it b}) correspond to
$(X_{\fe}^{\rm f}/X_{\fe \odot},\, X_{\oxy}^{\rm f}/X_{\oxy \odot})
=(0,\, 0)$ and
$(X_{\fe}^{\rm f}/X_{\fe \odot},\, X_{\oxy}^{\rm f}/X_{\oxy \odot})
=(0.1,\, 0.25)$, respectively
(same as Fig.\ \ref{fig2}{\it a} and {\it b}, respectively).
We see that Model A in Figure \ref{fig4}{\it b} seems to be the best of all the
models. However,
considering the uncertainty in the yields, we do not try any fine
tuning. The excess of the observed number of stars around
${\rm [Fe/H]}\sim 0.0$ is consistent with the data by
\citet{rocha00b}. As stated in \S~\ref{subsec:amr},
this may be due to the recent significant enrichment in the 
solar neighborhood.

\subsubsection{Evolution of [Fe/O]}

In order to test the effect of the limit-cycle behavior on the
[Fe/O] ratio, we examine the relation between [Fe/O]
and [Fe/H]. Since the oxygen is mainly produced by stars with
short lifetimes, the effect of the limit-cycle ISM is
reflected by the time evolution of the oxygen abundance.
On the other hand, the iron is also produced by
stars with long lifetimes and the information of the
oscillation of ISM phase is lost in the iron abundance.
Thus, we expect that [Fe/O] oscillates as the limit-cycle
evolution of ISM.

\begin{figure*}[htbp]
\figurenum{5}
\epsscale{1.1}
\plottwo{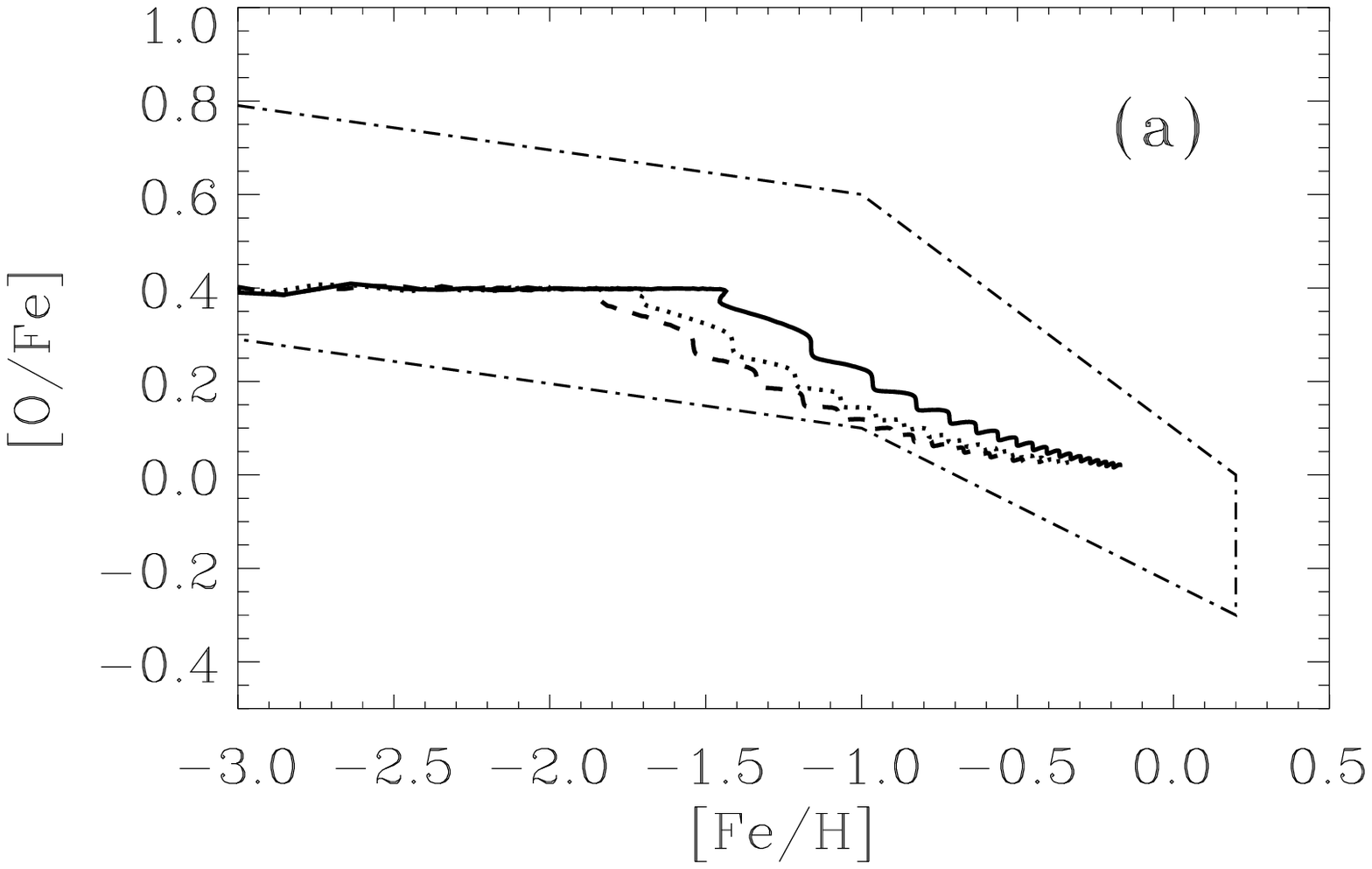}{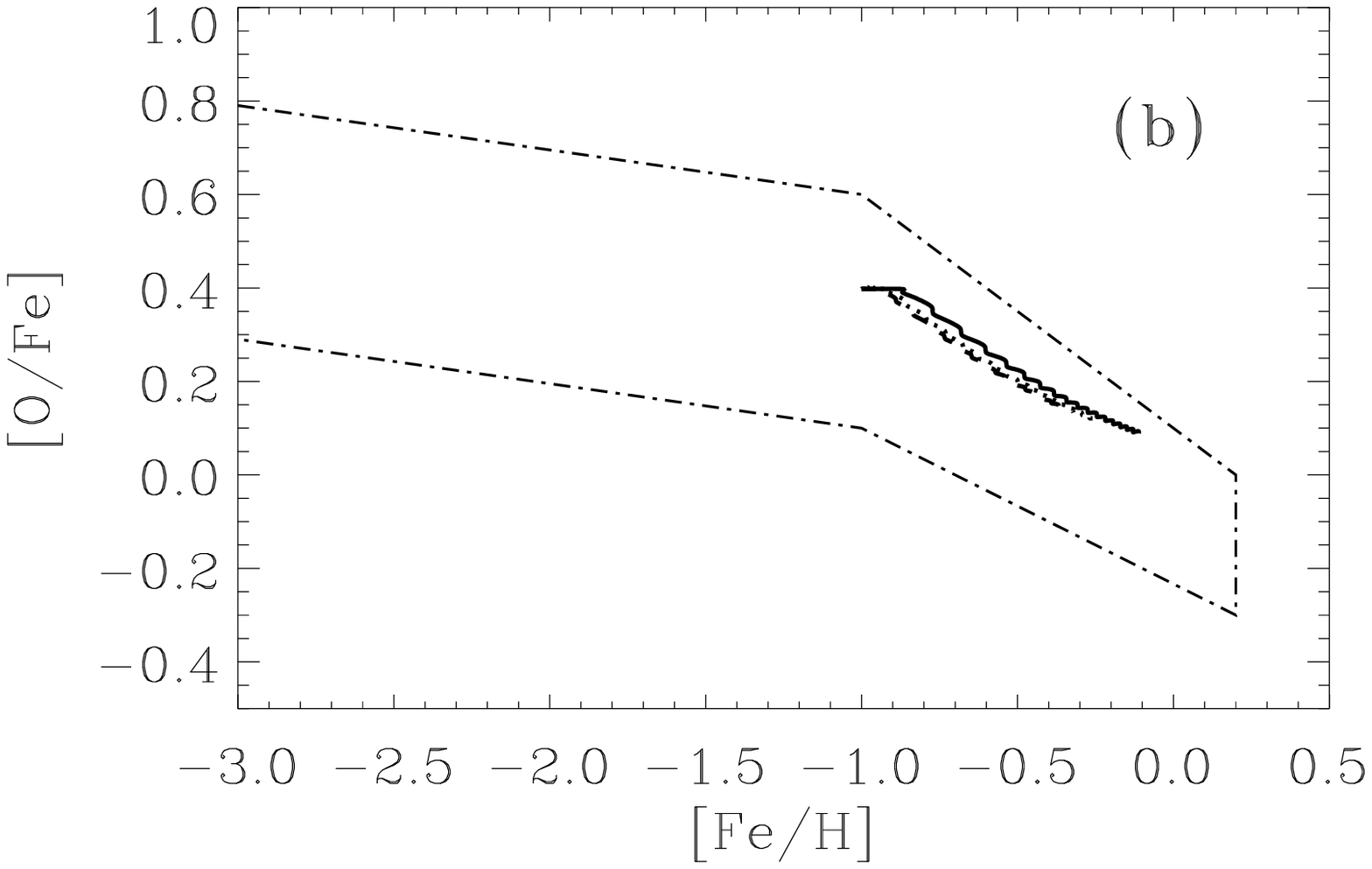}
\figcaption{Change in [O/Fe] against [Fe/H].
The solid, dotted, and dashed lines represent the results in
Models A, B, and C, respectively. The dot-dashed line
represents the observational data as summarized in Fig.\ 3
of \citet{kobayashi98}.
 ({\it a}) Abundance in the infalling gas is
assumed as ({\it a})
$(X_{\fe}^{\rm f}/X_{\fe \odot},\, X_{\oxy}^{\rm f}/X_{\oxy \odot})
=(0,\, 0)$ and ({\it b})
$(X_{\fe}^{\rm f}/X_{\fe \odot},\, X_{\oxy}^{\rm f}/X_{\oxy \odot})
=(0.1,\, 0.25)$.
\label{fig5}}
\end{figure*}

In Figure \ref{fig5}, we show [Fe/O]--[Fe/H] relation.
The solid, dotted, and dashed lines represent
the results in Models A, B, and C, respectively.
The two figures ({\it a}) and
({\it b}) correspond to
$(X_{\fe}^{\rm f}/X_{\fe \odot},\, X_{\oxy}^{\rm f}/X_{\oxy \odot})
=(0,\, 0)$ and
$(X_{\fe}^{\rm f}/X_{\fe \odot},\, X_{\oxy}^{\rm f}/X_{\oxy \odot})
=(0.1,\, 0.25)$, respectively
(same as Fig.\ \ref{fig2}{\it a} and {\it b}, respectively).
Indeed, we see that [Fe/O] oscillates. However, 
the amplitude of the oscillation is not large.
This is consistent with the age--metallicity relation
(Fig.\ \ref{fig2}). As mentioned in \S~\ref{subsec:amr},
the amount of metallicity reflects the past-integrated SFR;
thus, as long as the mass of newly formed stars is not
dominated in the total stellar mass, the present oscillation
has little influence on the metallicity evolution.

\section{DISCUSSIONS}\label{sec:discuss}

In this paper, we have modeled the oscillatory SFH proposed
observationally by R00. Our model is a
combination of an infall model developed in the field
of chemical evolution and the limit-cycle model proposed
by \citet{ikeuchi88} and his collaborators. We discuss
our result in the following two subsections.

\subsection{Limit-Cycle Star Formation History}

The oscillatory behavior of the Galactic SFH
proposed by R00 is modeled by using the limit-cycle model
of SFH. The limit-cycle behavior of the
three-phase ISM is suggested by \citet{ikeuchi88} and his
collaborators. Since the period of a limit-cycle orbit can be
$\sim 1$ Gyr within the framework of \citet{ikeuchi88},
the Galactic SFH is explained by 
the limit-cycle model of the ISM.

Recently, \citet{hirashita_kamaya00} have explained the observed
scatter of star formation activity of a sample of  spiral galaxies by
using the limit-cycle model. They provided a consistent modeling that
explains the variation in the scatter of star formation activity
among the morphological types (Sa--Sc) as shown in
Kennicutt et al.\ (1994).
Since the Galaxy is a spiral galaxy, an oscillatory SFH is
consistent with the variation of star formation 
activity seen in other spirals.

Kennicutt et al.\ (1994) have also presented the ratio of the present SFR
to the past averaged SFR (indicated as $b$ there).
{}From Figure 2 of R00, we see that $b$ (denoted as
$\SFR /\langle{\rm SFR}\rangle$ in 
R00) can be as large as 2--3 for the Galactic SFH. Since the
morphological type of the Galaxy is Sbc
(e.g., \citealt{binney98}, p.\ 171), $b=2$--3 is within the
range of the Sbc/Sc sample in Kennicutt et al.\ (1994, their
Fig.\ 6). This consistency implies that the oscillatory SFH
may be a common nature for all the spiral galaxies.

\subsection{Chemical Evolution in Limit-Cycle ISM}

The chemical evolution of the Galaxy has been investigated in the
framework of the limit-cycle ISM model.
We have found that the amplitude of the oscillatory behavior of the
metallicity is smaller than the observed scatter
(Figs.\ \ref{fig2} and \ref{fig5}). This indicates
that the observed scatter is not attributed to the
limit-cycle behavior. The scatter might be explained by
chemical inhomogeneity in the Galactic disk.

The oscillatory behavior of the metallicity is not prominent
because the metallicity reflects all the past SFH.
The integrated contribution from all the past SFH smoothes out the
oscillatory behavior of SFR.
Thus, from the viewpoint of chemical evolution we conclude that we
cannot distinguish between the ``smooth''
infall model without an oscillatory behavior and
the oscillatory infall model proposed in this paper.

Contrary to the metallicity, the dust-to-gas ratio can show
a oscillatory behavior \citep{hirashita00b}.
This is because the efficiency of the dust formation
changes according to phase changes in the gas. Moreover, dust
is efficiently destroyed when the mass fraction of the cold
gas is small.
This oscillation of dust amount may be important for
the evolution of infrared luminosity of galaxies.

\subsection{Another Possible Mechanism for the Variation of SFR}

\citet{rocha00c} gave some indication that the Magellanic Clouds
could play a role in the SFH of the Galaxy. It is meaningful
to explore their idea. Since our discussion is based on the
limit-cycle behavior inherent in the ISM as stated in \S~1,
we do not include the external force into our formulation.
Fortunately, \citet{ikeuchi83} have investigated the behavior of the
ISM in the presence of such an external force. Thus, we discuss
the influence of the Magellanic Clouds based on the
discussion in \citet{ikeuchi83}.

If a perturbation of the external force to the limit-cycle ISM
exists, both the period and the amplitude of the oscillation are
affected. Thus, it is not necessary that the period proper to the ISM
is about 1 Gyr. Even in the stable-focus case (\S~\ref{subsec:model}),
an oscillation emerges.
Furthermore, the qualitative behavior can be changed:
the system can show a chaotic orbit. Since the observational data
do not reject such a chaotic orbit affected by the perturbation
of the Magellanic Clouds, the strongly variable SFH of the Galaxy
might be due to the interaction with the
Magellanic Clouds. However, we note that the large scatter
of the star formation activity of the spiral sample
(e.g., Tomita et al.\ 1996) implies a general oscillatory behavior
of ISM in spiral galaxies. This is naturally explained if such an
oscillatory behavior is caused by the limit-cycle evolution
inherent in the ISM.

Finally, we would like to note that Chiappini et al.\ (1997) have
also considered SFR variation due to a different mechanism.
Their rapidly variable SFR is caused by a density
threshold for star formation: They assumed that
star formation occurs only if the surface density of the gas
exceeds a critical value, while we have assumed no threshold.
However, the timescale of the SFR variation in Chiappini et al.\
is much shorter than 1 Gyr. Although Chiappini et al.'s mechanism
may indeed present on a timescale much shorter than 1 Gyr, it is
necessary to introduce a mechanism different from Chiappini et al.\
in order to explain the SFR variability presented by 
R00. Thus, we have proposed a limit-cycle
scenario for the SFR in the Galactic disk.

\acknowledgments

We thank H. Rocha-Pinto, the referee, for invaluable comments
and suggestions that improved this paper very much.
We are grateful to S. Mineshige and H. Shibai for continuous
encouragement, H. Kamaya and A. Ibukiyama
for useful comments, and K. Yoshikawa for an excellent
computational environment.
H. H. was supported by the Research Fellowship of the Japan
Society for the Promotion of Science for Young Scientists.
A. B. acknowledges the hospitality of the department of astronomy at
Kyoto University where this project was started and financial support
from the Japan Society for the Promotion of Science.
We fully utilized the
NASA's Astrophysics Data System Abstract Service (ADS).

\appendix

\begin{figure*}[htbp]
\figurenum{B1}
\epsscale{0.6}
\plotone{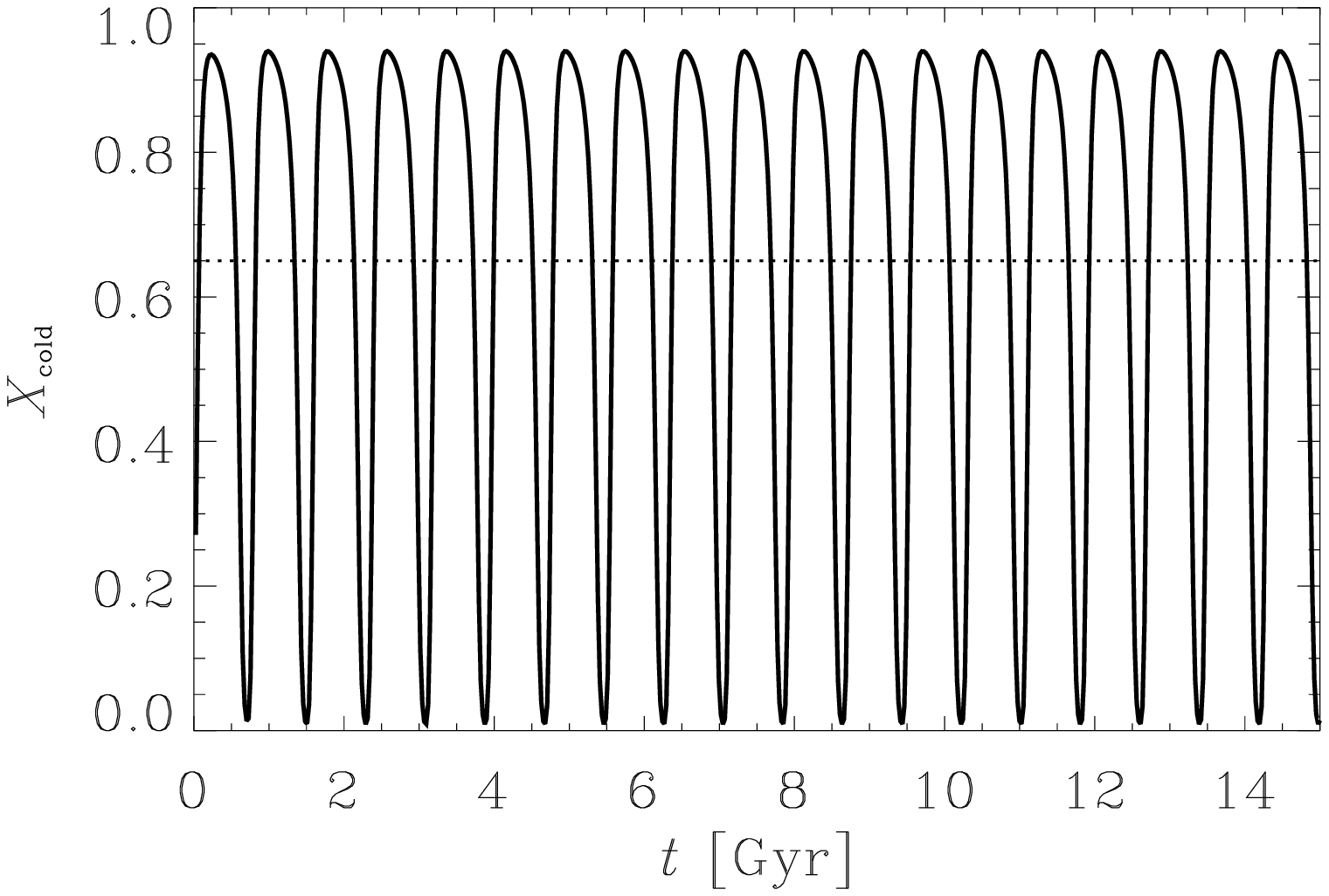}
\figcaption{Time evolution of the cold gas fraction
$X_{\rm cold}$ ({\it solid line}). The dotted line represents
the time-averaged value of $X_{\rm cold}$.
\label{figb1}}
\end{figure*}

\section{A. DETERMINATION OF $R_\ins $ AND $R_\del $}
\label{app:return}

$R_\ins $ and $R_\del $ are described as
\begin{eqnarray}
R_\ins  & = & \int_{m_{\rm l,\, ins}}^{m_{\rm u}}(m-w_m)\, \phi (m)\,
dm \, ,\\
R_\del  & = & \int_{m_{\rm l,\, del}}^{m_{\rm l,\, ins}}(m-w_m)\, \phi
(m)\, dm \, ,
\end{eqnarray}
where $\phi (m)$ is the initial mass function (IMF), which is normalized
so that the integral of $m\phi (m)$ in the full range of the stellar
mass (0.1--100$M_\odot$ in this paper) becomes unity; $m_{\rm u}$ is the 
upper mass cut-off of the stellar mass, and we here adopt
$m_{\rm u}=100M_\odot$; $m_{\rm l,\, ins}$ and
${m_{\rm l,\, del}}$ are set as $5M_\odot$ and $1M_\odot$,
corresponding to the stellar lifetime of $\tau$ (the parameter for
the delayed production) and the age of galaxies. If we adopt the
Salpeter's IMF ($\phi (m)\propto m^{-2.35}$), we obtain
$R_\ins =0.16$ and $R_\del =0.13$.

\section{B. DETERMINATION OF $\langle X_{\rm cold}\rangle$}

The cold gas mass fraction averaged over the Galactic lifetime,
 $\langle X_{\rm cold}\rangle$, is used in \S~\ref{subsec:param}.
It is estimated as follows.
First, the time evolution of the cold gas is calculated
based on the limit-cycle model in \S~\ref{sec:limit} by adopting the
parameters as $A=0.3$, $B=0.5$, and $1/c=10^7$ yr.
The time evolution of $X_{\rm cold}$ is shown in Figure \ref{figb1}.
Next, we estimate $\langle X_{\rm cold}\rangle$ by averaging
$X_{\rm cold}(t)$ over the Galactic age ($T_{\rm G}$) as
\begin{eqnarray}
\langle X_{\rm cold}\rangle =\frac{1}{T_{\rm G}}\int_0^{T_{\rm G}}
X_{\rm cold}(t)\, dt \, .
\end{eqnarray}


\end{document}